\documentclass[11pt,english]{article}
\usepackage{amsmath,latexsym,amssymb,amsthm,color}
\usepackage{fullpage}
\usepackage{ifthen,graphics,epsfig}
\usepackage{times}
\usepackage{graphicx,url}
\usepackage{babel}
\usepackage{graphicx} 
\usepackage{float}
\usepackage{epstopdf}
\usepackage{epsfig}
\usepackage{graphicx}
\usepackage{epstopdf}
\usepackage{graphicx}         
                           
\usepackage{amssymb}
\usepackage{epsfig}
\usepackage{babel}
\usepackage{epstopdf}
\usepackage[mathscr]{euscript}
\usepackage{amsmath,amsfonts,mathrsfs}
\usepackage{amsmath,amssymb}
\usepackage{calligra}
\usepackage{calrsfs}
\usepackage{here}
\usepackage{amsmath}
\usepackage{amsfonts}
\usepackage{amssymb}
\usepackage{diagbox}
\usepackage{makecell}
\usepackage{xspace}

\newcommand{\Xomit}[1]{}

\newcommand{\neighbors}{\mathit{neighbors}}
\newcommand{\palette}{\mathit{palette}}
\newcommand{\dtwocol}{\mathit{d2colors}}
\newcommand{\donecol}{\mathit{d1colors}}
\newcommand{\CLOCK}{\mathit{CLOCK}}

\newcommand{\df}[1]{}
\newcommand{\ccolor}{{\sc color}\xspace}
\newcommand{\term}{{\sc term}\xspace}

\newtheorem{theorem}{Theorem}
\newtheorem{lemma}{Lemma}

\newcommand{\toto}{xxx}

\newenvironment{proofL}{\noindent{\bf
Proof }} {\hspace*{\fill}$\Box_{Lemma~\ref{\toto}}$\par\vspace{3mm}}

\newcounter{linecounter}
\newcommand{\linenumbering}{\ifthenelse{\value{linecounter}<10}{(0\arabic{linecounter})}{(\arabic{linecounter})}}
\renewcommand{\line}[1]{\refstepcounter{linecounter}\label{#1}\linenumbering}
\newcommand{\resetline}[1]{\setcounter{linecounter}{0}#1}
\renewcommand{\thelinecounter}{\ifnum \value{linecounter} > 9\else 0\fi \arabic{linecounter}}

\newfloat{algorithm}{th}{lop}
\floatname{algorithm}{Algorithm}

\title{\bf  Optimal Collision/Conflict-free Distance-2 Coloring \\
            in  Synchronous  Broadcast/Receive  Tree Networks}

\author{Davide Frey$^{\dag}$~~  
        Hicham Lakhlef$^{\ddag}$~~ 
        Michel Raynal$^{\ddag,\star}$
~\\~\\
$^{\dag}$  INRIA Bretagne Atlantique, Rennes,  France \\
$^{\ddag}$  IRISA, Universit\'e de Rennes,  France \\
$^{\star}$  Institut Universitaire de France\\
{\small{\tt hicham.lakhlef@irisa.fr ~ raynal@irisa.fr}}
~\\~\\~\\
\centerline{Tech Report \# 2030 (19 pages), December 2015}\\
~\\
\centerline{IRISA, University of Rennes 1 (France)}
}
\date{}

\newpage

\begin{document}

\maketitle

\begin{abstract}
  This article is on message-passing systems where communication is
  (a) synchronous and (b) based on the ``broadcast/receive'' pair of
  communication operations.  ``Synchronous'' means that time is
  discrete and appears as a sequence of time slots (or rounds) such
  that each message is received in the very same round in which it is
  sent.  ``Broadcast/receive'' means that during a round a process can
  either broadcast a message to its neighbors or receive a message
  from one of them.  In such a communication model, no two neighbors
  of the same process, nor a process and any of its neighbors, must be
  allowed to broadcast during the same time slot (thereby preventing
  message collisions in the first case, and message conflicts in the
  second case).  From a graph theory point of view, the allocation of
  slots to processes is know as the distance-2 coloring problem: a
  color must be associated with each process (defining the time slots
  in which it will be allowed to broadcast) in such a way that any two
  processes at distance at most $2$ obtain different colors, while the
  total number of colors is ``as small as possible''.

  The paper presents a parallel message-passing distance-2 coloring
  algorithm suited to trees, whose roots are dynamically defined. This
  algorithm, which is itself collision-free and conflict-free, uses
  $\Delta+1$ colors where $\Delta$ is the maximal degree of the graph
  (hence the algorithm is color-optimal).  It does not require all
  processes to have different initial identities, and its time
  complexity is $O(d\Delta)$, where $d$ is the depth of the tree.  As
  far as we know, this is the first distributed distance-2 coloring
  algorithm designed for the broadcast/receive round-based
  communication model, which owns all the previous properties.

~\\~\\{\bf Keywords}: 
Broadcast/receive communication, Collision, Conflict, 
Distance-2 graph coloring, Message-passing, 
Network traversal, Synchronous system, Time slot assignment,
Tree network, Wireless network. 
\end{abstract}

\thispagestyle{empty}
\newpage
\setcounter{page}{1}

\section{Introduction}
Graph coloring is an important problem related to (optimal) resource
allocation, mainly used to establish an optimal order in which
resources have to be allocated to processes~\cite{DKR82,L81}.  The
distance-$k$ coloring problem consists in assigning colors to the
vertices of the graph such that no two vertices at distance at most
$k$ have the same color. Let us remember that minimum distance-$1$
vertex coloring is an NP-complete problem~\cite{GJ79}.

Coloring is (with leader election~\cite{A81} and
renaming~\cite{ABDPR90,CRR11}) one of the most important {\it symmetry
  breaking} problems encountered in distributed
computing~\cite{IRR10}. Solving such problems requires\df{I removed
  ``a seed''. It struck me as odd. But add it back if it's commonly
  used with this meaning.}  a pre-existing initial asymmetry from which
the problem can be solved. In a lot of cases, this initial asymmetry
is given by the assumption that no two processes have the same
identity.\footnote{Let us notice that one of the oldest
  symmetry-breaking problems is mutual exclusion, where the problem
  has not to be solved once for all, but repeatedly in a fair
  way~\cite{R86}.  See also the monograph~\cite{AE14} for
  impossibility results in distributed computing due to
  symmetry/indisguishability arguments.}

\paragraph{Distributed distance-1 coloring in the classical point-to-point
synchronous message-passing model}
Let us consider a distributed computing setting, where the processes
constitute the vertices of a graph, and the communication channels its
edges. The distributed distance-1 vertex coloring problem consists in
associating a color with each process such that (a) no two neighbors
have the same color, and (b) the total number of colors is as small as
possible. This process-coloring problem has essentially been
investigated in reliable synchronous networks where any two
neighboring processes are connected by a bi-directional channel on
which each of them can send and receive messages
(e.g.~\cite{P00,R10,R13,S07}).  The main results are described in the
monograph~\cite{BE14}.  

In these reliable point-to-point synchronous systems, processes
proceed in synchronized steps, usually called rounds. Each round
consists of three phases: during the first phase, each process sends
messages to its neighbors; during the second phase, each process
receives messages; and during the last phase, each process executes
local computation.  The fundamental synchrony property is that a
message is received in the very same round in which it is sent. Hence,
when solving a problem in this synchronous computation model, a
crucial attribute of a problem is the minimal number or rounds needed
to solve it.  As far as the distance-1 coloring problem is concerned,
it has been shown that, if the communication graph can be logically
oriented such that each process has only one predecessor (e.g., a tree
or a ring), $O(\log^*n)$ rounds are necessary and sufficient to color
the processes with three colors~\cite{CV86,L92} ($n$ being the total
number of processes\footnote{$\log^* n$ is the number of times the
  function $\log$ needs to be iteratively applied in $\log(\log(\log(
  ...(\log n))))$ to obtain a value $\leq 2$.  As an example, if $n$
  is the number of atoms in the universe, $\log^* n \backsimeq 5$.}).
Other d1-coloring algorithms are described in several articles
(e.g.~\cite{BE11,BEK14,GPS88,KW06}).  They differ in the number of
rounds they need and in the number of colors they use to implement
d1-coloring. Both the algorithms in~\cite{BE11,BEK14} color the
vertices with $(\Delta+1)$ colors; the first one requires
$O(\Delta+\log^* n)$ rounds, while the second one uses $O(\log
\Delta)$ rounds.  An algorithm described in~\cite{GPS88} is for tree
graphs or graphs where each vertex has two neighbors; it uses three
colors and $O(\log^* n)$ rounds. Another algorithm presented in the
same paper addresses constant-degree graphs; it uses $(\Delta+1)$
colors and $O(\log^* n)$ rounds. The algorithm presented
in~\cite{KW06} requires $O(\Delta \log \Delta +\log^* n)$ rounds.
These algorithms assume that the processes (vertices) have distinct
identities, which are their initial colors. They proceed iteratively,
each round reducing the total number of colors.

\paragraph{Distance-2 and distance-3 coloring in shared memory and
  message-passing  models}
A first class of algorithms addresses the distance-2 vertex coloring
problems in systems where communication is through shared memory,
message-passing, or a mix of the
two~\cite{BCGMBO05,BGMBC08,GMP02}. These algorithms find their
motivation in the application of distance-2 coloring to scientific
computing. As a result, they do not fit well the characteristics of
wireless networks.

On the other hand, wireless protocols apply or require distance-2 and
distance-3 coloring algorithms to prevent packet
collisions~\cite{chipara11:conflict}. To this end, they build a TDMA
(Time Division Multiple Access) schedule~\cite{R97} from the result of
the coloring process.  TDMA allows processes to share the same
frequency channel by dividing the signal into different time slots,
one per color. Hence, protocols based on distance-2 coloring guarantee
that nodes can transmit messages undisturbed during their time
slots. Those based on distance-3 coloring additionally allow receiving
nodes to acknowledge unicast messages in the sender's slot.


With this motivation, some authors have proposed self-stabilizing
algorithms to solve the distance-2
problem~\cite{BM09,gairing-distanceTwo} For example,~\cite{BM09}
presents a self-stabilizing distance-2 algorithm that uses a constant
number of variables on each node and that stabilizes in $O(\Delta^2
m)$ moves and uses at most $\Delta^2$ colors, where $m$ is the number
of edges. But these algorithms do not take the broadcast nature of the
wireless medium into account, and their operation may thus results in
significant packet collisions. 

Differently, CADCA~\cite{jemili13:collision} is a distributed distance-2
coloring algorithm that takes into account the risk of collisions
during the coloring process. To limit this risk, CADCA organizes the
nodes to be colored in concentric layers around a sink node. The
coloring process proceeds in three phases: the first colors layers 1,
4, 7, ...; the second colors layers 2, 5, 8, ...., and the third
colors layers 3, 6, 9. Each such phase uses a specific color palette
to avoid conflicts between nodes in different layers and exploits five
stages in which the nodes in a layer select colors and resolve the
conflicts that arise during the process. However, the algorithm
requires nodes to know their position with respect to the sink node,
and most importantly it does not entirely eliminate packet collisions,
it only reduces their number.

With respect to the distance-3 version of a problem,~\cite{HT04}
proposes a self-stabilizing algorithm. Processes compute a maximal
independent set and then use it to assign themselves colors. The
self-stabilizing part of the algorithm gathers information about each
process's 3-hop neighborhood, and does generate collisions. However,
the protocol uses a special TDMA slot to continuously run the
self-stabilizing protocol without interfering with colored slots.
Serena~\cite{serena} uses a similar approach and also presents a
distance-3 coloring algorithm in a broadcast-receive model. However,
it does not  use a special time-slot to limit the impact of the
collisions occurring during the coloring process.  The protocols we
present in this paper, on the other hand, do not lead to any conflict
or collision and do not need any special time slots.



\paragraph{Content of the paper}
Differently from the previous articles, we propose a collision- and
conflict-free algorithm that solves the distance-2 (d2) coloring problem in
synchronous networks where (a) processes communicate by broadcasting
and receiving messages, and (b) collisions and conflicts are not
prevented by the communication model.  A collision occurs when a
process receives messages from two or more neighbors in the same
round. A conflict occurs when, during the same round, two neighbors
send a message to each other.

In this broadcast/receive communication model (which covers practical
system deployments), there is not a dedicated communication medium for
each pair of processes, but a single shared communication medium for
each pair composed of a process and all its neighbors.  Examples of
such communication media are encountered in wireless networks such as
sensor networks. In such networks, collision-freedom and
conflict-freedom do not come for free, and the algorithms built on top
of them must be collision/conflict-free to ensure the consistency of
the messages that are exchanged, and consequently the progress of
upper-layer applications.

This paper is on collision/conflict-free d2-coloring for the
synchronous broadcast/receive communication model, where the processes
are connected by a tree network.  Considering such a context, it
presents an algorithm which uses $(\Delta+1)$ colors, and is
consequently optimal with respect to the number of colors. This
algorithm relies on two assumptions to break symmetry: (a) it assumes
that a process (not predetermined in advance) receives an external
message that defines it as the root of the tree; (b) it assumes that
any two processes at distance less 
than or equal to $2$ have distinct identities\footnote{Let us notice
  that this assumption states nothing more than the fact that a
  process is able to distinguish its neighbors based on their
  identities.} (hence, depending on the structure of the tree, lots of
processes can have the same identity). Its round complexity is
$O(d\Delta)$ where $d$ is the depth of the tree. Moreover, no process
needs to know $n$, $\Delta$, or the depth of the tree. Hence a process
has no information on the global structure of the tree.  Its initial
knowledge is purely local: it is restricted to its identity, and the
identities of its neighbors.


\paragraph{Roadmap}
The paper consists of~\ref{sec:conclusion} sections.
Section~\ref{sec:model} presents the synchronous broadcast/receive
model, and Section~\ref{sec:D2-coloring} introduces the distance-2
coloring problem.  Then, two distributed (message-passing) distance-2
coloring algorithms suited to trees are presented.  The presentation
is incremental.  Section~\ref{sec:sequential-d2-tree-algorithm}
presents first a simple distributed distance-2 coloring algorithm
which exploits a sequential tree traversal algorithm as a skeleton, on
which are appropriately grafted statements implementing distance-2 the
coloring.  Then, Section~\ref{sec:parallel-d2-tree-algorithm} presents
a distributed distance-2 coloring algorithm based on a parallel
traversal of the tree.  This second algorithm extends the basic
coloring principles introduced in the first algorithm.

\section{Synchronous Broadcast/Receive Model}
\label{sec:model}


\paragraph{Processes, initial knowledge, and the communication graph}
The system model consists of $n$ sequential processes denoted 
$p_1$, ..., $p_n$, connected by a tree communication network. 

Each process $p_i$ has an identity $id_i$, which is known only by
itself and its neighbors (processes at distance $1$ from it).  The
constant $\neighbors_i$ is a local set, known only by $p_i$, including
the identities of its neighbors (and only them).  As noticed in the
Introduction, in order for a process $p_i$ not to confuse its
neighbors, it is assumed that no two processes at distance less
than or equal to $2$ have distinct identities.  Hence, any two
processes at distance greater than $2$ may have the same
identity. When computing bit complexities, we will assume that any
process identity is encoded in $\log_2 n$ bits.

Let $\Delta_i$ denote the degree of a process $p_i$
(i.e. $|\neighbors_i|$) and let $\Delta$ denote the maximal degree of
the process graph ($\max\{\Delta_1,\cdots,\Delta_n\}$).  While each
process $p_i$ knows $\Delta_i$, no process knows $\Delta$ (a process
$p_x$ such that $\Delta_x=\Delta$ does not know that $\Delta_x$ is
$\Delta$).

When considering a process $p_i$, $1 \leq i \leq n$, the integer $i$ 
is called its index. Indexes are not known by the processes. 
They are only a notation convenience used as a subscript 
to distinguish processes and their local variables.

\paragraph{Timing model}
We assume that processing durations are equal to $0$. This is
justified by the following observations: (a) the duration of the local
computations of a process is negligible with respect to message
transfer delays, and (b) the processing duration of a message may be
considered as a part of its transfer delay.

Communication is synchronous in the sense that there is an upper bound
$D$ on message transfer delays, and this bound is known by all the
processes (global knowledge). From an algorithm design point of view,
we consider that there is a global clock, denoted $\CLOCK$, which is
increased by $1$, after each period of $D$ physical time units. 
Each value of $\CLOCK$ defines what is usually called a {\it time slot}
or a {\it round}.

\paragraph{Communication operations}
The processes are provided with two operations denoted ${\sf
  broadcast}()$ and ${\sf receive}()$.  A process $p_i$ invokes ${\sf
  broadcast}$ {\sc tag}$(m)$ to send the message $m$, whose type is
{\sc tag}, to all its neighbors.  It is assumed that a process invokes
${\sf broadcast}()$ only at a beginning of a time slot.  When a
message {\sc tag}$(m)$ arrives at a process $p_i$, this process is
immediately warned of it, which triggers the execution of operation
${\sf receive}()$ to obtain the message. Hence, a message is always
received and processed during the time slot --round-- in which it was
broadcast.

From a linguistic point of of view, we use the two following
{\bf when} notations when writing algorithms, where ${\sf predicate}$ 
is a predicate involving $\CLOCK$ and possibly local variables of the
concerned process.  
\begin{tabbing} 
$~~~~~~~~~~~~~~~~~~~~~$
{\bf  when} {\sc  tag}$(m)$ {\bf  is received do} processing of the message. \\
$~~~~~~~~~~~~~~~~~~~~~$
{\bf  when} ${\sf predicate}$ 
                         {\bf do} code entailing at most one ${\sf broadcast}()$
invocation. 
\end{tabbing}

\paragraph{Message collision and message conflict}
Traditional wired round-based synchronous systems assume a dedicated a
communication medium for each pair of processes (i.e., this medium is
not accessible to the other processes).  Hence, in these systems a
process $p_i$ obeys the following sequential pattern during each
round: (a) first $p_i$ sends a message to all or a subset of its
neighbors, (b) then $p_i$ receives the messages sent to it by its
neighbors during the current round, and (c) finally executes a local
computation which depends on its local state at the beginning of the
round and the messages it has received during the current round.

The situation is different in systems such as wireless networks (e.g.,
sensor networks), which lack a dedicated communication medium per pair
of processes.  A process $p_i$ shares a single communication medium
with all its neighbors, and ``message clash'' problems can occur,
each message corrupting the other ones, and being corrupted by
them. Consider a process $p_i$, these problems are the following.
\begin{itemize}
\vspace{-0.1cm}
\item 
If two  neighbors of $p_i$ invoke the operation ${\sf broadcast}()$  
during the same time slot (round), a message {\it collision} occurs.  
\vspace{-0.2cm}
\item 
If $p_i$ and one  of its neighbors 
invoke ${\sf broadcast}()$  during the same time slot 
(round), a message {\it conflict} occurs.  
\end{itemize}
As already indicated, this paper considers this broadcast/receive
communication model. This implies that protocols must prevent
collisions and conflicts to ensure both message consistency and
computation progress.

\section{The Distance-2 Tree Coloring Problem}
\label{sec:D2-coloring}

\paragraph{Solving the collision/conflict problem}
To prevent collisions and conflicts involving a process $p_i$, only a
single process in the set $\{p_i\}\cup \neighbors_i$ can obtain the
right to communicate during a given round. To this end, we associate
each process with time slots (rounds) in which it can broadcast a
message, while none of its 2-hop neighbors can broadcast during these
time slots. When considering the whole set of processes, this
assignment must be optimal in terms of numbers of colors (ideally
allowing as many processes as possible to broadcast during the same
round).

This problem is a well-known graph coloring problem called 
{\it distance-2} coloring. The aim is to design distributed algorithms
associating a color with each process (which will define the 
time-slots during which it  will be allowed to broadcast) such 
that the following properties are satisfied.

\paragraph{Definition}
\begin{itemize}
\vspace{-0.2cm}
\item Validity: 
The final color of each process belongs to $\{0,..., \Delta\}$. 
\vspace{-0.2cm}
\item Consistency: 
No two processes at distance $\leq 2$ have the same color.  
\vspace{-0.2cm}
\item Termination: Each process obtains a color and one process knows that 
this occurred. 
\end{itemize}
Let us observe that, as at least one process has $\Delta$ neighbors, 
$\Delta+1$ different  colors are necessary.  The Validity property
states that we are looking for {\it distributed} algorithms which ensure
that $\Delta+1$ is also a tight upper bound. As we will see such algorithms 
exist for tree networks.  

\paragraph{Using the colors to define the time slots}
The colors obtained by the processes are used as follows, where
$color_i$ is the color obtained by process $p_i$.  The time slots
(rounds) during which $p_i$ is allowed to broadcast a message to its
neighbors correspond to the values of $\CLOCK$ such that $\big((\CLOCK
\mbox{ mod } (\Delta +1)\big)= color_i)$.  As we will see, these time
slots are different from the time slots used during the (sequential
and parallel) distributed distance-2 algorithms which are presented
below. It follows that these algorithms must provide each process with
the (initially unknown) value of $\Delta$.

\section{Sequential Distance-2  Coloring of a Tree}
\label{sec:sequential-d2-tree-algorithm}
This section presents a distributed distance-2 coloring algorithm in
which there are neither message collisions, nor message conflicts.
This algorithm is sequential in the sense that its skeleton is a
depth-first tree traversal in which the control flow (implemented by
appropriate messages) moves sequentially from a process to another one. 

\subsection{A  sequential algorithm}
Algorithm~\ref{fig:DFTree-traversal-algorithm} assumes that a single
process receives a message, {\sc start}$()$, which defines it as the
root of the tree. (As noticed in the Introduction, this introduces the
initial asymmetry\df{removed 'seed' from here too} needed to solve the
symmetry-breaking problem we are interested in.)  This external
message causes the receiving process, $p_r$, to simulate the reception
of a fictitious message, {\sc color}$(id_r,id_r,-1,\emptyset)$.  This
message initiates a depth-first traversal of the tree.

\paragraph{Messages}
The algorithm uses two types of messages: {\sc color}$()$ and {\sc
  term}$()$.  As each message is broadcast by its sender and received
by all its neighbors, it carries the identity of its destination
process.  Hence, when a process receives a message $m$, it discards
$m$ if it is not the destination of $m$ (predicate $dest \neq id_i$ at
line~\ref{Seq-04} and line~\ref{Seq-19}).

These messages implement a depth-first traversal of the tree
network~\cite{R13}. Each carries the identity of its destination
$(dest)$, the identity of its sender $(sender)$, and the color of its
sender $(sender\_cl)$. A message {\sc color}$()$ additionally carries
the colors of the already colored neighbors of the sender
$(\donecol)$.

\paragraph{Local variables}
Each process $p_i$ manages the following local variables. 
\begin{itemize}
\vspace{-0.1cm}
\item $state_i$ (initialized to $0$) is used by $p_i$ to manage the
  progress of the tree traversal.  Each process traverses five
  different states during the execution of the algorithm. States $1$
  and $3$ are active: a process in state $1$ sends a \ccolor$()$ message to
  a child, while a process in state $3$ sends a \term$()$  message to its
  parent.  States $0$ and $2$ are waiting states. Nodes listen on
  the broadcast channel but cannot send any message. Finally, state
  $4$ identifies local termination. 
\vspace{-0.2cm}
\item $parent_i$ saves the identity of the process $p_j$ from which
  $p_i$ received the message {\sc color}$(id_i,-,-,-)$; $p_i$ receives
  exactly one such  message.  This process, $p_j$, defines the parent
  of $p_i$ in the tree.  The root $p_r$ of the tree, defined by the
  reception of the external message {\sc start}$()$, is the only
  process such that $parent_r = id_r$.
\vspace{-0.2cm}
\item $sender\_cl_i$ records the color of the parent of $p_i$. $p_i$ 
  receives this information in the parent's \ccolor message.\df{This
    was collapsed into the previous one. I found it nicer to split
    them since they are two different variables albeit related.}
\vspace{-0.2cm}
\item $\donecol_i$ is a set containing the colors of the neighbors of $p_i$,
that have  already obtained their color. 
\vspace{-0.2cm}
\item $to\_color_i$ (initialized to $\neighbors_i$) is a set
  containing the identities of the neighbors of $p_i$ not yet
  colored. 
\vspace{-0.2cm}
\item  $color_i$ contains the color of $p_i$. \df{Same as above. This
    was collapsed into the previous but I preferred to split them.}
\end{itemize}

\begin{algorithm}
\centering{
\fbox{
\begin{minipage}[h]{150mm}
\footnotesize
\renewcommand{\baselinestretch}{2.5}
\resetline
\begin{tabbing}
aaaaA\=aA\=aaA\=aaA\=aaA\=aaA\kill

{\bf  Initialization:}
    $state_i\leftarrow 0$; $to\_color_i \leftarrow \neighbors_i$.\\~\\

\line{Seq-01}  \> 
{\bf  when} {\sc  start}$()$ {\bf  is received do} 
\%  a single process $p_i$  receives this external message  \% \\

\line{Seq-02}  
\> \> $p_i$ executes the lines~\ref{Seq-04}-\ref{Seq-09} as if it received 
      the message    {\sc color}$(id_i,id_i,-1,\emptyset)$.\\~\\

\line{Seq-03}  
{\bf when} {\sc color}$(dest,sender,sender\_cl,\donecol)$ 
     {\bf  is received  do} \\

\line{Seq-04}  \>\> 
{\bf if}      ($dest\neq id_i)$ 
     {\bf then} discard the message
           (do not execute lines~\ref{Seq-05}-\ref{Seq-09}) {\bf end if};\\

\line{Seq-05}  \>\>  
$parent_i \leftarrow sender$;
$sender\_cl_i \leftarrow  sender\_cl$;
$\donecol_i \leftarrow \{sender\_cl\}$;\\

\line{Seq-06}  \>\>  
$\palette \leftarrow$ sequence $0,1,2,...$ 
without the colors in  $\{sender\_cl_i\}\cup \donecol$;\\

\line{Seq-07}  \>\> 
$color_i   \leftarrow$ first color in $\palette$;\\

\line{Seq-08}  \>\>  
$to\_color_i \leftarrow \neighbors_i\setminus\{parent_i\} $;\\

\line{Seq-09}  \>\> 
{\bf if}  $(to\_color_i \neq  \emptyset)$
{\bf then} $state_i \leftarrow 1$ {\bf else} $state_i \leftarrow 3$ 
{\bf end if}.\\~\\

\line{Seq-10}  \> {\bf when} 
($(\CLOCK$ increases$)$ $\wedge (state_i\in \{1,3\})\big)$ {\bf do}\\

\line{Seq-11}  \>\> {\bf if} \=  $(state_i=1)$\\

\line{Seq-12}  \>\>\> {\bf then} \=   
  $next \leftarrow$ any $id_k\in to\_color_i$;\\

\line{Seq-13}  \>\>\>\>
   ${\sf broadcast}$ {\sc color}$(next,id_i,color_i,\donecol_i)$;
   $state_i \leftarrow 2$ \\

\line{Seq-14}  \>\>\> {\bf else} \>
{\bf if} \=  $(parent_i=id_i)$ \= {\bf then} \=
the root $p_i$ claims termination \\

\line{Seq-15}  \>\>\>\>\>\>{\bf else} \>
 ${\sf broadcast}$ {\sc term}$(parent_i,id_i,color_i)$\\

\line{Seq-16}  \>\>\>\> {\bf end if}; $state_i\leftarrow 4$ \\

\line{Seq-17} \>\>  {\bf end if}. \\~\\


\line{Seq-18} \>
{\bf when} {\sc term}$(dest,id,sender\_cl)$ {\bf is received} {\bf do} \\
 \> \% the tree rooted at process $id$ is properly colored \% \\

\line{Seq-19} \>\>
{\bf if} $(dest \neq id_i)$ 
   {\bf then} discard the message
      (do not execute lines~\ref{Seq-20}-\ref{Seq-21}) {\bf end if};\\

\line{Seq-20} \>\> 
$to\_color_i \leftarrow  to\_color_i \setminus \{id\} $;
$\donecol_i \leftarrow \donecol_i \cup \{sender\_cl\}$;\\

\line{Seq-21}  \>\> 
{\bf if} $(to\_color_i \neq  \emptyset)$ 
 {\bf then} $state_i \leftarrow  1$ 
 {\bf else} $state_i \leftarrow  3$ 
{\bf end if}.

\end{tabbing}
\normalsize
\end{minipage}
}
\caption{Distributed depth-first-based distance-2 coloring of a tree  
(code for $p_i$)}
\label{fig:DFTree-traversal-algorithm}
}
\end{algorithm}

\paragraph{Description of the algorithm}
Nodes start the algorithm in state $0$, waiting for a {\sc
  color}$(id_i,-,-,-)$ message.  A process, $p_i$, receives such a
message exactly once.  When it receives it, it is visited for the
first time by the depth-first tree traversal.  It consequently assigns
the values of the message parameters to its local variables
$parent_i$, $sender\_cl_i$ and $\donecol_i$ (line~\ref{Seq-05}).
Then, it computes its color, which is different from the color of the
sender of the message and from the colors of the sender's already
colored neighbors (lines~\ref{Seq-06}-\ref{Seq-07}). Finally, $p_i$
updates $to\_color_i$, and transitions to a new state: $1$ if it has
any children, or $3$ if it is a leaf node. This prepares the progress
of the tree traversal, which will take place at the next time slot
(round) (lines~\ref{Seq-08}-\ref{Seq-09}).

When $p_i$ enters the new time slot with $state_i\in\{1,3\}$
(line~\ref{Seq-10}), it operates as follows to ensure the progress
of the tree traversal.
\begin{itemize}
\vspace{-0.1cm}
\item If $state_i=1$, it means that $p_i$ has neighbors that have not
  yet been colored. In this case, $p_i$ selects one of them, and makes
  the tree traversal progress by broadcasting a message, {\sc
    color}$(next,id_i,color_i,\donecol_i)$, which will be processed
  only by $next$ (line~\ref{Seq-13}).  Then, $p_i$ moves into
  $state_i=2$ and waits until it receive a \term message, {\sc
    term}$(id_i,next,-)$.

\vspace{-0.1cm}
\item If $state_i=3$, it means that all the neighbors of $p_i$ have
  been colored. In this case, if $p_i$ is the root, the distance-2
  coloring has terminated (line~\ref{Seq-14}). Otherwise, if $p_i$ is
  not the root, it broadcasts the message {\sc
    term}$(parent_i,id_i,color_i)$ to inform its parent that the
  sub-tree of which it is the root has been colored
  (line~\ref{Seq-15}).  In both cases, $p_i$ transitions to
  $state_i=4$, thereby indicating that the algorithm is terminated as
  far as $p_i$ is concerned (local termination).
\end{itemize}

Finally, when $p_i$ receives the message {\sc
  term}$(id_i,id,sender\_cl)$, it first updates its local variables
$to\_color_i$ and $\donecol_i$ according to the received values
(line~\ref{Seq-20}). Then, it updates $state_i$ according to the value
of $to\_color_i$ (indicating whether it has colored all its neighbors,
line~\ref{Seq-21}). In the first case, it moves to state $1$ and
continues traversing another sub-tree. Otherwise it moves to state
$3$, which will then evolve into state $4$ (signaling local
termination) as indicated above.

\paragraph{How a process learns the value of $\Delta$}
A process maintains a local variable $max\_d_i$ initialized to
$\Delta_i$, and each message {\sc term}$()$ now carries this value.
When a process $p_i$ receives a message {\sc term}$(max\_d,-,-,-)$
it executes the update statement $max\_d_i\leftarrow \max(max\_d_i,max\_d)$.  
Finally, when the root process $p_r$ claims termination, it launches a second 
traversal of the tree to inform the other processes.  We do not describe such 
a propagation of the value of $\Delta$ here. This will be done in
Section~\ref{sec:inform-others}  in the context of a parallel tree traversal.

\subsection{Proof and cost of the algorithm}

\begin{lemma}
\label{collision-conflict-freedom}
Algorithm~{\em\ref{fig:DFTree-traversal-algorithm}} is 
both collision-free and conflict-free. 
\end{lemma}

\begin{proofL}
Let us first observe that, due to the assignment of the variable 
$state_i$ at line~\ref{Seq-09} (in the processing of a message 
{\sc color}$()$),  or line~\ref{Seq-21}  (in the processing of a message 
{\sc term}$()$), a process $p_i$ is allowed to  broadcast one and only 
one message  when it executes line~\ref{Seq-13} or~\ref{Seq-15}.
It follows from the associated assignment of the control value $2$ or $4$ to 
$state_i$ (line~\ref{Seq-13} or~\ref{Seq-15}), that $p_i$  cannot broadcast 
other messages {\sc color}$()$ or {\sc term}$()$  before executing
line~\ref{Seq-21} (i.e., before it receives a message {\sc term}$()$).

Let us now observe that, due to line~\ref{Seq-04}, a message 
{\sc color}$()$ broadcast by a process at line~\ref{Seq-13}
is processed by a single destination  process. 
Hence, the control flow generated by these messages remains sequential, moving 
sequentially from a parent process to a child process. 
Similarly, the  control flow generated by the messages {\sc term}$()$ 
(broadcast at line~\ref{Seq-15}) moves sequentially from a child  process $p_i$
to its  parent process (whose identity is saved in $parent_i$).

The collision-freedom and conflict-freedom properties of the algorithm follow
directly from the sequentiality of the control flow realized by the messages
{\sc color}$()$ and {\sc term}$()$.
\renewcommand{\toto}{collision-conflict-freedom}
\end{proofL}

\begin{lemma}
\label{lemma-three-properties}
Algorithm~{\em\ref{fig:DFTree-traversal-algorithm}} satisfies the 
Validity, Consistency, and Termination properties. 
\end{lemma}

\begin{proofL}
  Let us first prove the following claim. \\
  Claim. For any $p_i$ and  at any time, $|\donecol_i|<\Delta$. \\
  Proof of the claim.  Let us consider the local variable $\donecol_i$
  of a process $p_i$.  This variable is initialized at
  line~\ref{Seq-05}, when $p_i$ receives a message {\sc color}$()$ for
  the first time. It then contains the color of its parent in the
  tree. When the algorithm progresses, the color of a child $p_j$ of
  $p_i$ is added to $\donecol_i$ (line~\ref{Seq-20}) when $p_i$
  receives from $p_j$ a message {\sc term}$()$ carrying $p_j$'s color.
  It follows that, when $p_i$ issues its last broadcast of {\sc
    color}$(child,-,-,\donecol_i)$, $child$ is the identity of its
  only child without a color, and $\donecol_i$ contains the colors of
  all the other neighbors of $p_i$. hence, $|\donecol_i|<\Delta$. End
  of the proof of the claim.

The Validity property follows from the following observation.
When a process $p_i$ selects its color, it follows from the previous claim
and   lines~\ref{Seq-06}-\ref{Seq-07} that
$|\{parent\_cl\} \cup \donecol|\leq \Delta$. Consequently, 
there is at least one free color in the set $\{0,\cdots,\Delta\}$. 

To prove the Proper-Coloring property let us first observe that, due
to (a) the initialization of $\donecol_j$ done when a process $p_j$
receives a message {\sc color}$()$ from its parent
(line~\ref{Seq-05}), and (b) the updates that follow when it receives
messages {\sc term}$()$ from its children (line~\ref{Seq-20}), it
follows that $\donecol_j$ contains the colors of the already colored
neighbors of $p_j$.  Hence, when a process $p_i$ selects a color
(lines~\ref{Seq-06}-\ref{Seq-07}), the colors of the already colored
processes at distance 2 from it are in the set $\donecol$ carried by
the message {\sc color}$()$ entailing $p_i$'s coloring.  Due to
line~\ref{Seq-06}, $p_i$ does not select any of these colors.

The Termination property follows from the termination of the sequential 
traversal, at the end of which the root learns the algorithm has terminated. 
\renewcommand{\toto}{lemma-three-properties}
\end{proofL}

\noindent
The following theorem is an immediate consequence of the previous lemmas. 
\begin{theorem}
\label{theo:seq-traversal-coloring}
Algorithm~{\em\ref{fig:DFTree-traversal-algorithm}} 
is a collision-free and conflict-free distance-2 coloring algorithm for trees. 
\end{theorem}

\paragraph{Cost of the algorithm}
(Let us recall that a process identity can be encoded with $O(\log n)$ bits.)
There are two message types. A message {\sc term}$()$ carries two 
process identities and a color. 
A message {\sc color}$()$ carries two process identities, 
a color, and set of at most $(\Delta-1)$ colors. 
It follows that a message carries at most $2\log n + \Delta \log \Delta$ bits.

A tree of maximal degree $\Delta$ has at least $x\geq \Delta-1$ leaves.  
Let us first count the number of broadcasts of a message  {\sc color}$()$. 
The root issues at most $\Delta$  broadcasts; 
a process, which is neither the root nor a leaf, issues
at most $(\Delta-1)$  broadcasts; and a leaf  issues no broadcast
of such a message. It follows that there are 
$\Delta + (n-1-x)(\Delta-1)$  broadcasts of a message  
{\sc color}$()$. As $x\geq \Delta-1$, the number of broadcasts of 
a message {\sc color}$()$  is upper bounded by $
\Delta + (n-\Delta)(\Delta-1)$.  
 
Each process which is not the root of the tree  issues exactly one 
broadcast  of a message {\sc term}$()$. It follows that there are $(n-1)$
broadcasts  of such a message.

\section{Parallel Distance-2  Coloring of a Tree}
\label{sec:parallel-d2-tree-algorithm}

This section presents a distributed distance-2 coloring for trees,
which based on a parallel traversal of the tree network, with feedback
(i.e., the dynamically defined root process that launches the network
traversal learns of the end of the traversal).  This algorithm can be
seen as improvement of the previous algorithm in terms of time
efficiency.


\subsection{A parallel algorithm}
\paragraph{Underlying principle}
Let us first observe that any two children of a process must be prevented 
from  broadcasting simultaneously a message to their neighbors. 
This is because, being issued by processes at  distance at most two, 
such simultaneous broadcasts will create a collision at least at the
parent process.

The idea to prevent this vicinity/concurrency problem is first to
direct a parent process to compute the colors of its children, and
then, as soon as a process has obtained a color, to allow it to broadcast 
a message ({\sc color}$()$ or {\sc term}$()$) only during the time slots
(rounds) associated with its color.  To this end, each process uses
the values provided by the global clock ($\CLOCK$).

\paragraph{Messages and local variables}
The message types implementing the parallel tree traversal are the
same as in Algorithm~\ref{fig:DFTree-traversal-algorithm}.  Similarly,
the local variables $parent_i$, $to\_color_i$, $color_i$,
$sender\_cl_i$, and the constant $\neighbors_i$, have the same meaning
as in Algorithm~\ref{fig:DFTree-traversal-algorithm}.

Each process $p_i$ manages an additional variable $nb\_cl\_parent_i$,
initially $0$, which will contain the number of colors needed to color
its parent $p_j$ and its neighbors (processes of $\neighbors_j$). This
value is known to $p_i$ when it receives its first message {\sc
  color}$()$ (which defines its sender $p_j$ as $p_i$'s parent). Each
process has also a constant $nb\_cl_i=\Delta_i+1$ which represents the
number of colors needed to color itself and its neighbors.

\begin{algorithm}[ht]
\centering{
\fbox{
\begin{minipage}[t]{150mm}
\footnotesize
\renewcommand{\baselinestretch}{2.5}
\resetline
\begin{tabbing}
aaaA\=aaaA\=aaA\=aaA\=aaA\=aaA\kill

{\bf  Initialization:} $nb\_cl_i= \Delta_i+1$; 
    $state_i\leftarrow 0$; $max\_nb\_cl_i \leftarrow nb\_cl_i$.\\~\\

\line{P-01}  \> 
{\bf  w}\={\bf hen} {\sc  start}$()$ {\bf  is received do} 
      \%  a single process $p_i$ receives this external message  \% \\

\line{P-02}  \> \> 
$p_i$ executes lines~\ref{P-04}-\ref{P-08} as if it received the message 
   {\sc color}$(\{\langle id_i,cl \rangle\}, id_i, -1, nb\_cl_i)$ \\
\>\>  where $cl= (\CLOCK+1) \mbox{ mod } nb\_cl_i$.\\~\\

\line{P-03}  \> 
{\bf  when} {\sc  color}$(\mathit{pairs}, sender, sender\_cl, nb\_cl\_parent)$ 
{\bf  is received} {\bf do} \\

\line{P-04}  \>\> 
{\bf if}  ({\sc color}$()$ already received) 
     {\bf then} discard the message,
           do not execute lines~\ref{P-05}-\ref{P-08}) {\bf end if};\\    
      
\line{P-05}  \>\> 
       $parent_i \leftarrow sender$;
       $to\_color_i  \leftarrow \mathit{neighbors_i}\setminus\{sender\} $;
       $sender\_cl_i \leftarrow sender\_cl$;\\

\line{P-06}  \>\> 
$color_i\leftarrow cl$ such that  
       $\langle id_i,cl \rangle \in \mathit{pairs}$;\\
       
       
\line{P-07}  \>\> $nb\_cl\_parent_i \leftarrow nb\_cl\_parent$;   \\    

\line{P-08}  \>\> {\bf if} 
$to\_color_i\neq \emptyset$  
   {\bf then}  $state_i \leftarrow 1$ {\bf else}  $state_i \leftarrow 3$ 
   {\bf end if}.\\~\\

\line{P-09}  \> {\bf when} 
$\big((\CLOCK \mbox{ mod } nb\_cl\_parent_i) = color_i) 
                                 \wedge (state_i\in \{1,3\})\big)$ {\bf do}\\

\line{P-10}  \>\>
 {\bf if} \=  $(state_i=1)$\\

\line{P-11}  \>\>\> {\bf then} \= 
$pairs\_for\_children_i \leftarrow$ empty set of pairs;\\

\line{P-12}  \>\>\>\>
$palette \leftarrow$ sequence $0,1,...$ 
without the colors  $color_i$ and $sender\_cl_i$;\\

\line{P-13}  \>\>\>\>
{\bf for} \={\bf each} \= $k\in \mathit{to\_color_i}$ {\bf do}  \\

\line{P-14}  \>\>\>\>\>\>
$cl \leftarrow$ first color in $palette$; \\

\line{P-15}  \>\>\>\>\>\>
suppress $cl$ from $palette$
and add $\langle k,cl \rangle$ to  $pairs\_for\_children_i$\\

\line{P-16}  \>\>\>\> {\bf end for};\\

\line{P-17}  \>\>\>\>
 ${\sf broadcast}$ 
{\sc color}$(pairs\_for\_children_i,id_i, color_i, nb\_cl_i)$; 
     $state_i\leftarrow 2$ \\

\line{P-18} \>\>\> {\bf else} \> 

 ${\sf broadcast}$ {\sc term}$(parent_i,id_i)$;  
 $state_i\leftarrow 4$   \% $p_i$ and its neighbors are colored \% \\

\line{P-19} \>\>  {\bf end if}. \\~\\

\line{P-20}  \>
{\bf when}  {\sc term}$(dest,id)$ {\bf is received} {\bf do} \\

\line{P-21} \>\> {\bf if} $(dest \neq id_i)$ 
   {\bf then} discard the message
      (do not execute lines~\ref{P-22}-\ref{P-25}) {\bf end if};\\

\line{P-22} \>\> $to\_color_i  \leftarrow to\_color_i  \setminus\{ id\}$;  \\

\line{P-23}   \>\> {\bf if} \= ($to\_color_i= \emptyset$)  \\

\line{P-24}  \>\>\> {\bf then if}  ($parent_i=id_i$) 
 {\bf then}  the root $p_i$ claims termination 
 {\bf else}   $state_i\leftarrow 3$  {\bf end if} \\

\line{P-25}  \>\> {\bf end if}.

\end{tabbing}
\normalsize
\end{minipage}
}
\caption{Parallel distributed distance-2 coloring of a tree  (code for $p_i$)}
\label{fig:tree-parallel-coloring-algorithm}
}
\end{algorithm}

\paragraph{Description of the algorithm}

To simplify the presentation, we assume that the initial value of
$\CLOCK$ is $-1$.  As in the sequential version, a single process $p_r$
receives the external message {\sc start}$()$, that defines it as the
root of the tree. Moreover, this reception entails the fictitious sending 
of the message {\sc color}$(\{\langle id_r,cl \rangle\}, id_r, -1, nb\_cl_r)$
where $cl= (\CLOCK+1) \mbox{ mod } nb\_cl_r$ (line~\ref{P-02}).
Similarly to Algorithm~\ref{fig:DFTree-traversal-algorithm}, 
according to the values carried by the message {\sc color}$()$, 
$p_r$ initializes its local variables and obtains the color $cl=1$
(lines~\ref{P-05}-\ref{P-07}).  It finally updates $state_r$.

Then, when $\CLOCK=1$, $p_r$ executes lines~\ref{P-10}-\ref{P-19}.
More generally, these lines are executed by any process $p_i$ as soon
as, after it received its first message {\sc color}$()$, the color it
obtained ($color_i$) is such that $color_i= \CLOCK \mbox{ mod }
nb\_cl\_parent_i$ (line~\ref{P-09}). Hence, as the algorithm (a)
allows a process to broadcast only at a round corresponding to its
color, (b) allows only colored processes to broadcast {\sc color}$()$
or {\sc term}$()$ messages, and (c) ensures the distance-2 coloring
property (see the proof) without any message collision or message
conflict during its execution.

Let us consider a process $p_i$ such that the predicate of line~\ref{P-09} 
is satisfied. There are two cases. 
\begin{itemize}
\vspace{-0.1cm}
\item If $state_i=1$, the neighbors of $p_i$ are not colored. 
Process $p_i$ computes then a color for each of its children 
(lines~\ref{P-11}-\ref{P-16}), and broadcasts a message {\sc color}$()$ 
to inform them of it (line~\ref{P-17}). It is easy to see that the messages 
{\sc color}$()$ implements a parallel traversal of the tree from the 
root to the leaves. Then,  $p_i$ enters a waiting period by 
setting $state_i$ to $2$. 
\vspace{-0.2cm}
\item If $state_i=3$, all neighbors of $p_i$ (and all processes in their 
sub-trees) are colored. In this case, $p_i$ forwards this information to 
its parent. As its local participation to the algorithm is terminated,  
$p_i$ sets $state_i$ to $4$. 
\end{itemize}

Finally, when a process $p_i$ receives a message {\sc term}$()$ from
one of its children, it first updates  $to\_color_i$ accordingly
(line~\ref{P-22}). If this set is empty, it claims termination if it
is the root. Otherwise, it assigns $3$ to $state_i$, so that it will
inform its parent of its local termination at the first time slot
(round) at which its color satisfies the predicate of line~\ref{P-09}.

\subsection{Proof  and cost of the algorithm}

\begin{lemma}
\label{lemma:par-noCollision-noConflict}
Algorithm~{\em\ref{fig:tree-parallel-coloring-algorithm}}
is collision-free and conflict free. 
\end{lemma}

\begin{proofL}
Let us observe that a process $p_i$ is allowed to broadcast a message 
{\sc color}$()$ (line~\ref{P-16}) only if 
$\big((\CLOCK \mbox{ mod } cl\_bound) = color_i) \wedge (state_i=1)\big)$. 
As such a sending entails the assignment 
$(state_i\leftarrow 2$ (and $state_i$ never decreases), 
it follows that $p_i$ issues at most one such broadcast.  
The same occurs for the broadcast of a message {\sc term}$()$. 

Due to the initialization of its local variable $state_i$, 
no process $p_i$ can broadcast a message until one of them receives the message 
${\sc start}()$. Let  us assume that (without loss of generality) that 
$\CLOCK=0$ when a process a process $p_i$ receives this message. 
It follows from the text of lines~\ref{P-01}-\ref{P-08} that, before
$\CLOCK$ increases, we have 
$color_i=1$ and $state_i=1$ (if the graph is not a singleton), or 
$color_i=1$ and $state_i=3$ (if the graph is a singleton).  

Hence, when $\CLOCK=1$, $p_i$  executes lines~\ref{P-09}-\ref{P-19}
and broadcasts a message {\sc color}$()$ (line~\ref{P-17}), or a message
{\sc term}$()$ (line~\ref{P-18}),
while no other process can broadcast a message. Moreover, 
due to the color assignment done by $p_i$ at lines~\ref{P-12}-\ref{P-16}, 
its neighbors obtain different colors  
when they receive the message {\sc color}$()$ from $p_i$ at time $\CLOCK=1$. 

Due to the color computation done by $p_i$ at line~\ref{P-12}-\ref{P-16} 
(when  $\CLOCK=1$), no two of its neighbors have the same color, and none
of them has the same color as $p_i$. It follows from the time predicate 
of line~\ref{P-09} that no two processes of the set $\neighbors_i \cup id_i$  
can broadcast during the same time slot. 

Let $p_j$ be a neighbor of the root process $p_i$. 
It follows from line\ref{P-12} that, when $p_j$ selects colors for 
its neighbors, it assigns them colors which are different among themselves
and different from its own color and the color of $p_i$. 
Hence, due the  time predicate of line~\ref{P-09}, it follows 
that no two processes  of the set $\neighbors_j \cup \{id_j\}$ can broadcast 
during the same time slot.   
The same  reasoning applies to the neighbors of $p_j$, etc., which completes 
the proof of the lemma. 
\renewcommand{\toto}{lemma:par-noCollision-noConflict}
\end{proofL}

\begin{lemma}
\label{lemma:color-size}
Let $p_i$ and  $p_j$ be two processes such that such  $parent_i=id_j$. 
The color of $p_i$ belong to the set $\{0, ..., \Delta_j\}$. 
\end{lemma}

\begin{proofL}
  Let $p_i$ be any children pf $p_j$.  The color of $p_i$ is defined
  by $p_j$ when it executes the lines~\ref{P-12}-\ref{P-16}.  The
  local palette of $p_j$ is then the sequence $0,1,...$, from which
  its own color ($color_j$) and the color of its neighbor which is
  parent ($sender\_cl$) are suppressed (line~\ref{P-12}).  Hence, at
  most two integers (one in case of the root) are suppressed from the
  first $\Delta_j+1$ non-negative integers from which the palette is
  built.  Hence, $p_j$ can color its neighbors with up to $\Delta_j-1$
  ($\Delta_j$ if $p_j$ is the root) colors with values less than
  $\Delta_j+1$ and different from that of its parent.
  \renewcommand{\toto}{lemma:color-size}
\end{proofL}

\begin{lemma}
\label{lemma:par-proper-color-bound}
Algorithm~{\em\ref{fig:tree-parallel-coloring-algorithm}} 
uses at most $\Delta+1$ different colors to the processes, and no
two processes at distance $\leq 2$ have the same color.
\end{lemma}

\begin{proofL}
Let us consider a process $p_i$. It has at most $\Delta$ neighbors. 
It follows from lines~\ref{P-12}-\ref{P-16}, that all its  
neighbors (including the process from which it received the message 
{\sc color}$()$) are  assigned different colors, and those are 
different from its color $color_i$. Hence no two processes at distance 
$\leq 2$ have the same color. As $\Delta=\max(\Delta_1,\cdots,\Delta_n)$, 
it follows from  Lemma~\ref{lemma:color-size} that at most 
$(\Delta+1)$ colors are used by the algorithm. 
\renewcommand{\toto}{lemma:par-proper-color-bound}
\end{proofL}

\begin{lemma}
\label{lemma:par-termination}
Any process is colored, and (only after all processes are colored) 
this is known by the root process. 
\end{lemma}

\begin{proofL}
Once a process received an external message {\sc start}$()$, 
it becomes the root of the tree, it takes a color, 
and broadcasts a message {\sc color}$()$
to its neighbors at line~\ref{P-17} as soon as the
predicate of line~\ref{P-09}  becomes satisfied. 
Then, each of these neighbor processes  broadcasts a 
message {\sc color}$()$ to their neighbors, etc. 

As soon as its neighbors are  colored, a process $p_i$ is such 
that $state_i=3$, and consequently it broadcasts a message 
{\sc term}$()$  at line~\ref{P-18}. 
Let  $p_j$ be the parent process that  broadcasts the message {\sc color}$()$ 
received by $p_i$. When $p_j$ has received message {\sc term}$()$ 
from all its children, it will proceed to $state_j=3$ (line~\ref{P-24}) 
and will broadcast {\sc term}$()$  at line~\ref{P-18} as soon as the 
time predicate of line~\ref{P-09} becomes satisfied. 

It follows from the previous observations that the process that received 
the message {\sc start}$()$ eventually claims termination at line~\ref{P-24}. 
\renewcommand{\toto}{lemma:par-termination}
\end{proofL}

\noindent
The following theorem is an immediate consequence of the previous lemmas. 
\begin{theorem}
\label{theo:parallel-coloring}
Algorithm~{\em\ref{fig:tree-parallel-coloring-algorithm}}
is a collision-free and conflict-free distance-2 coloring 
parallel algorithm for trees. 
\end{theorem}

\paragraph{Cost of the algorithm}
Each process which is not a leaf issues one broadcast of a message 
{\sc color}$()$, and each process which is not the root issues one 
broadcast of a message {\sc term}$()$. Let $x$ be the number of leaves.
There are consequently $(2n-(x+1))$ broadcasts.  As $x\geq\Delta -1$, 
the number of broadcasts is upper bounded by $2n-\Delta$.

As far as time complexity is concerned, we have the following.  Let
$d$ be the depth of the tree, and assume $\CLOCK=0$ when a process
$p_r$ receives the message {\sc start}$()$, which defines it as the 
root of the tree. It follows from their color assignment 
(lines~\ref{P-12}-\ref{P-16}) that $p_r$'s 
children start one after the other at times $1$, $2$, ..., $\Delta_r$.
Let $p_x$ be the child of $p_r$ that obtains the color $\Delta_r$.
It broadcasts a message  {\sc color}$()$ to its own children at time 
 $\Delta_r+1$, and its child with the highest color will 
do the same at time  $\Delta_r+\Delta_i$. Etc. 
As this worst pattern can repeat along a path of the tree, 
it follows than  process does not receive a message {\sc color}$()$ 
before time  $d\Delta$. 

An analogous reasoning applies to the ``return'' messages {\sc term}$()$ 
starting from the leaves to the root. Hence, the time complexity is 
$O(d\Delta)$. (Let us remind that $d=O(\log_\Delta n)$
when the tree is well-balanced.

\subsection{Informing  processes of termination}
\label{sec:inform-others}

\paragraph{Global vs local termination}
Algorithm~\ref{fig:tree-parallel-coloring-algorithm} implements a
parallel distance-2 coloring, but only the root learns that the
algorithm has terminated (global termination). A non-root process
$p_i$ knows only that the sub-tree of which it is the root has terminated
(local termination).  This section, enriches this algorithm so that any
process learns about global termination.

\paragraph{The extended algorithm}
This extended algorithm is made up of
Algorithm~\ref{fig:tree-parallel-coloring-algorithm} where 
line~\ref{P-18} is modified, and the
statement ``{\bf when} {\sc term}$(id)$ {\bf is received do} ...''
(lines~\ref{P-20}-\ref{P-24}) is replaced by the statements described in 
Algorithm~\ref{fig:-tree-parallel-coloring-termination}.  Moreover,
each process $p_i$ manages an additional local variable denoted
$max\_nb\_cl_i$, which is initialized to $\Delta_i+1$.

When considering base Algorithm~\ref{fig:tree-parallel-coloring-algorithm} 
and its extension Algorithm~\ref{fig:-tree-parallel-coloring-termination}, 
the lines with the same number are the same in both algorithms, 
the lines suffixed by a ``prime'' are modified lines, and the lines 
N1-N7 are new lines. 

\begin{algorithm}[ht]
\centering{
\fbox{
\begin{minipage}[t]{150mm}
\footnotesize
\renewcommand{\baselinestretch}{2.5}
\begin{tabbing}
aaaaA\=A\=aaA\=aaA\=aaA\=aaA\kill


(\ref{P-20}')  \>
{\bf when}  {\sc term}$(dest,id, max\_nb\_cl)$ {\bf is received} {\bf do}\\

(\ref{P-21}) \>\> {\bf if} $(dest \neq id_i)$ 
   {\bf then} discard the message
      (do not execute lines~\ref{P-22}-\ref{P-25}) {\bf end if};\\

(\ref{P-22}) \>\> $to\_color_i \leftarrow to\_color_i \setminus\{ id\}$;\\

(N1) \>\> 
    $max\_nb\_cl_i \leftarrow \max(max\_nb\_cl_i, max\_nb\_cl);$  \\
  
(\ref{P-23})   \>\> {\bf if} \= ($to\_color_i= \emptyset$)  \\

(\ref{P-24}')  \>\>\> {\bf then if}  ($parent_i=id_i$) 
 {\bf then}   $state_i\leftarrow 5$
 {\bf else}   $state_i\leftarrow 3$  {\bf end if} \\

(\ref{P-25})  \>\> {\bf end if}.\\~\\


(N2) \> {\bf when}  
$\big((\CLOCK \mbox{ mod } max\_nb\_cl_i) = color_i) 
                                 \wedge (state_i =5)\big)$ {\bf do}\\

(N3)  \>\>

 {\bf if}  $(|\neighbors_i|\neq 1)$  {\bf then} 
 ${\sf broadcast}$ {\sc end}$(id_i,\mathit{max\_nb\_cl_i})$ {\bf end if}; 
 $state_i\leftarrow 6$.\\~\\


(N4) \> {\bf when}  {\sc end}$(parent,max\_cl)$ {\bf is received} {\bf do} \\

(N5)  \>\> 
  {\bf if}  $(parent=parent_i)\wedge (state_i=4)$ \\

(N6)  \>\>\> 
 {\bf then} $max\_nb\_cl_i \leftarrow \max(max\_nb\_cl_i, max\_cl)$;
  $state_i\leftarrow 5$ \\

(N7) \> \> {\bf end if}. 
 
\end{tabbing}
\normalsize
\end{minipage}
}
\caption{Parallel distance-2 coloring of a tree: propagation of the termination}
\label{fig:-tree-parallel-coloring-termination}
}
\end{algorithm}

The message {\sc term}$(id_i)$ broadcast by a process $p_i$ at
line~\ref{P-18} must now carry the current value of $max\_nb\_cl_i$,
i.e., $p_i$ broadcasts {\sc term}$(parent_i,id_i,max\_nb\_cl_i)$.  The local
variable $max\_nb\_cl_i$ of a process $p_i$ is updated at line N1. In this way,
starting from the leaves, these local variables allow the root to know
the value $(\Delta+1)$ (upper bound on the number of colors needed by
a process to color itself and its neighbors).

When the root learns about global termination, it proceeds to
$state_i=5$ (line~\ref{P-24}'). At this point, the value of its local
variable $max\_nb\_cl_i$ is $\Delta+1$. Moreover, its new local state
$state_i$ allows it to inform its children about global termination by
broadcasting the message {\sc end}$()$ carrying the value $\Delta+1$
(lines~N2-N3).

Finally, when a process $p_i$, which is not the root (hence
$state_i=4$) receives an {\sc end} $(parent,max\_cl)$ message, from
its parent, it updates $max\_nb\_cl_i$ and proceeds to $state_i=5$
(lines N4-N7), which allows it to forward the {\sc end} message to its
children, if any (lines N2-N3).  Let us notice that, as the children
of a process $p_x$ do not broadcast {\sc end}$()$ messages during the
same time slot (round), there is no collision of these messages at
their parent. But since the parent simply discards these messages, a
trivial optimization may consist in relaxing the collision-freedom
constraint. Specifically, the children could send their {\sc end}()
messages in parallel as their parent does not need to receive
them. Independently of the optimization, it follows from the
propagation of the {\sc end}$()$ messages that, eventually, all the
processes are such that $state_i=6$. When this occurs, they learn about
global termination.

\paragraph{Remark}
Let us consider the situation where, while the coloring algorithm has 
terminated (i.e., each  process has obtained a color and knows $\Delta$),  
a new process $p_x$ wants to enter the tree and obtains a color. 
Let $p_i$ be the process  chosen to be $p_x$'s parent. 
If $\Delta_i = \Delta$, $p_i$ cannot accept $p_x$ as a child
(this would require an additional color).  Differently, 
if $\Delta_i<\Delta$, it is easy to dynamically add $p_x$ as a child of $p_i$. 
To this end, $p_i$ takes the first color $cl$ in $0$, $1$, ..., $\Delta$, 
which is different from its own color and the colors of its neighbors.  
Then, during its next time slot, $p_i$  broadcasts the  message 
{\sc new}$(cl,\Delta)$, which, when received by $p_x$, gives it a proper 
color. Let us notice that this ``join'' does not require $p_x$ to have an 
identity.  It is consequently possible for $p_i$ to also assign to $p_x$  an 
identity not belonging to $\{id_i\} \cup \neighbors_i$.

\subsection{Merging two trees with the same maximal degree}
Considering two trees $T_1$ and $T_2$, which have the same maximal 
degree $\Delta$, let  $p_x$  be a process of $T_1$ and  
$p_y$ a process of $T_2$,  both having a degree less than $\Delta$. 

It is easy to see that, if $\big(id_x\notin (\{id_y\} \cup
\neighbors_y)\big)\wedge \big(id_y\notin (\{id_x\} \cup 
\neighbors_x)\big)$, the trees $T_1$ and $T_2$ can be ``added'' to  
compose a single tree made  up of $T_1$ and $T_2$ connected by the
additional edge $(p_x,p_y)$, This composition preserves the colors
previously assigned by two  independent executions of the algorithm, the 
one which assigned colors to $T_1$ and the one which assigned colors to $T_2$.

\section{Conclusion}
\label{sec:conclusion}
Synchronous networks where the time is decomposed in a sequence of
time slots (rounds) and the communication operations are ``broadcast a
message to neighbors'' and ``receive a message a neighbor'', are prone
to message collisions (which occur when two neighbors of a process
send it a message during the same time slot), and message conflicts
((which occur when a process and one of its neighbors broadcast during
the same time slot).  Distance-2 coloring solves this problem by
assigning a color to each process such that there is a matching of
time slots with colors which prevents message collisions/conflicts
from occurring.

This paper has presented a distributed algorithm which solves the
distance-2 coloring problem in tree networks.  This algorithm is based
on a parallel tree traversal algorithm skeleton on which are grafted
appropriate coloring assignments.  It is itself
collision/conflict-free. It uses only $\Delta+1$ colors ($\Delta$
being the maximal degree of the network), which is optimal. Its time
complexity is $O(d\Delta)$ (where $d$ is the depth of the tree). This
algorithm does not require a process to initially know more than its
identity and the ones of its neighbors. Moreover, any two processes at
distance greater than $2$ are not prevented from having the same
identity (which is important for scalability issues).  Let us also
notice that this algorithm is  relatively simple (a first-class property).

A very challenging issue is now the design of a parallel 
collision/conflict-free distance-2 coloring algorithm for synchronous
broadcast/receive systems whose communication graph is more general
than a tree. As, when considering sequential computing, 
distance-2 coloring is an NP-complete problem, the design of such a
distributed algorithm using a ``reasonable'' number of colors does not seem
to be a  ``trivial''  challenge. (A sequential algorithm suited to an 
arbitrary graph is presented in Appendix~\ref{sec:seq-arbitrary-graph}.) 


\section*{Acknowledgments}
This work has been partially supported by the Franco-Hong Kong ANR-RGC
Joint Research Programme 12-IS02-004-02 CO$^2$2Dim, the Franco-German
DFG-ANR Project 40300781 DISCMAT (devoted to connections between
mathematics and distributed computing), the French ANR project
SocioPlug (ANR-13-INFR-0003), and the Labex CominLabs excellence
laboratory (ANR-10-LABX-07-01) through the DeSceNt project.



\appendix 
\section{A sequential distributed algorithm for an arbitrary graph}
\label{sec:seq-arbitrary-graph}

Algorithm~\ref{fig:coloring-of-an-arbitrary-graph} is a sequential 
distance-2 coloring algorithm 
for an  arbitrary (connected) graph. The design of this algorithm 
is the same the one of  Algorithm~\ref{fig:DFTree-traversal-algorithm}.
It added complexity comes from the fact it allows the sequential control flow
(implemented by the messages {\sc color}$()$ and {\sc term}$()$
in Algorithm~\ref{fig:DFTree-traversal-algorithm})
to back track when a process discovers  a  coloring conflict.  
This backtracking is implemented by the messages  {\sc correct}$()$,
{\sc corrected\_color}$()$, and {\sc resume\_coloring}$()$.

\begin{algorithm}[t]
\centering{
\fbox{
\begin{minipage}[t]{150mm}
\footnotesize
\renewcommand{\baselinestretch}{2.5}
\resetline
\begin{tabbing}
aaaaA\=aA\=aaA\=aaA\=aaA\=aaA\kill

{\bf  init}: $\donecol_i \leftarrow \emptyset$; $sender_i \leftarrow 0$; 
             $\dtwocol_i \leftarrow \emptyset$; $state_i \leftarrow 0$; 
             $to\_color_i \leftarrow \neighbors_i$.\\~\\

{\bf  when} {\sc  start}$()$ {\bf  is received do} \\ 

\line{AG-01}  
\>  the reception of this message 
    defines its receiver $p_i$  as the root of the tree;\\
\>  this process $p_i$ simulates then  the sending of   
    {\sc color}$(id_i,id_i,-1, 0,\emptyset)$ to itself.\\~\\


{\bf when} {\sc color}$(dest,sender,sender\_cl, proposed\_color, \donecol)$ 
     {\bf  is received  do} \\

\line{AG-02} \> 
$to\_color_i \leftarrow to\_color_i \setminus \{sender\}$;

 $\dtwocol_i \leftarrow \dtwocol_i \cup \donecol$;\\

\line{AG-03} \> {\bf if} \= 
$ (dest=id_i) \wedge ((sender\_cl \in \donecol_i) \vee (proposed\_color \in  
(\donecol_i\cup\dtwocol_i))) $  \\

\line{AG-04} \>\> {\bf then} \= 
  $state_i \leftarrow 1$; $sender_i \leftarrow sender$;\\

\line{AG-05}  \>\> {\bf else}  \> 
{\bf if} \= $(dest = id_i) \wedge (to\_color_i \neq \emptyset) $  \\

\line{AG-06} \>\>\>\> {\bf then} \=  $state_i \leftarrow 2$; 
$color_i\leftarrow proposed\_color$;  $parent_i \leftarrow sender$; 
$\donecol_i \leftarrow \donecol_i \cup \{sender\_cl\}$;\\

\line{AG-07}  \>\>\>\> {\bf else} \> 
{\bf if} \= $(dest = id_i) \wedge (to\_color_i = \emptyset) $  

 {\bf then} \= $state_i \leftarrow 3$;  $parent_i \leftarrow sender$; 
  {\bf end if} \\

\line{AG-08} \>\>\>   {\bf end if} \\

\line{AG-9} \>   {\bf end if};\\

\line{AG-10} \>
{\bf if} $(dest \neq  id_i)$ 
   {\bf then} $\dtwocol_i \leftarrow \dtwocol_i \cup \{proposed\_color\}$;
  $\donecol_i \leftarrow \donecol_i \cup \{sender\_cl\}$  {\bf end if}. \\~\\

  
{\bf when} $(CLOCK \, $increases$)$  \\
  
\line{AG-11} \> {\bf if} \= $ (state_i=1) $  

 {\bf then} \=   
$color_i\leftarrow$ the first color in $0, 1...$ which is not in
  $\donecol_i \cup \dtwocol_i$;\\

\line{AG-13}\>\>\>
${\sf broadcast}$ {\sc correct}$(sender_i, id_i, color_i, \donecol_i)$;  
$state_i \leftarrow 0 $  \\

\line{AG-14}\>{\bf end if}; \\
  
\line{AG-15} \> {\bf if} \= $ (state_i=2) $  

 {\bf then} \=    

 $color\_for\_child \leftarrow$ the first color in $0,1..$  which is not in
  $\donecol_i \cup color_i$;\\

\line{AG-17}  \>\>\>\>\>\>
 $next \leftarrow$ any $id_k\in to\_color_i$;\\

\line{AG-18}  \>\>\>\>\>\>
        ${\sf broadcast}$ 
  {\sc color}$(next,id_i, color_i, color\_for\_child, \donecol_i); 
state_i \leftarrow 0 $ \\

\line{AG-19}\>{\bf end if}; \\
  
\line{AG-20} \> {\bf if}  $ (state_i=3) $ {\bf then} \= 
        ${\sf broadcast}$ 
{\sc term}$(parent_i, color_i, id_i)$; $state_i \leftarrow 0 $  {\bf end if}; \\

   \line{AG-21} \> {\bf if}  $ (state_i=4) $ {\bf then} \=
${\sf broadcast}$ {\sc corrected\_color }$(sender_i, parent_i, id_i, color_i)$;
   $state_i \leftarrow 0 $ {\bf end if}; \\
  
 \line{AG-22} \> {\bf if}  $ (state_i=5) $ {\bf then} \= 
       ${\sf broadcast}$ {\sc resume\_coloring }$(sender_i,  id_i)$; 
       $state_i \leftarrow 0 $ {\bf end if};   \\
  
  \line{AG-23} \> {\bf if}  $ (state_i=6) $ {\bf then} \= 

${\sf broadcast}$ {\sc corrected\_color }$(-1, -1, id_i, corrected\_cl_i)$ ;
 $state_i \leftarrow 0 $ {\bf end if}.    \\~\\


{\bf when} 
{\sc term}$(dest, id, color)$ {\bf is received} {\bf do} \\

\line{AG-24} \>
{\bf if} $(dest \neq  id_i)$ 
   {\bf then} discard the message
      (do not execute lines~\ref{AG-25}-\ref{AG-30}) {\bf end if};\\

\line{AG-25}  \> 
$to\_color_i \leftarrow  to\_color_i \setminus \{id\} $;\\

\line{AG-26}  \> 
{\bf if} $(to\_color_i =  \emptyset)$ 
  \% the neighbors of $p_i$  are properly colored \% \\

\line{AG-27}  \>  \> {\bf then} \= 
{\bf if} $(parent_i=id_i)$ \= {\bf then}  
                 \= the root claims the graph is colored  {\bf else}    
    $state_i \leftarrow 3$ 
 {\bf end if}\\

\line{AG-28}  \>\> {\bf else} \>
$\donecol_i \leftarrow \donecol_i\cup \{color\}$;

 $next \leftarrow$ any $id_k\in to\_color_i$;\\

\line{AG-29}  \>\>\> 
 $color\_to\_child \leftarrow$ the first color in $0,1..$  which is not in
  $\donecol_i \cup color_i$; $state_i \leftarrow 2$ \\

\line{AG-30}  \>  {\bf end if}. \\~\\

{\bf when} 
{\sc correct}$(dest,sender, color, \donecol)$ {\bf is received} {\bf do} \\

\line{AG-31} \>
{\bf if} $(dest \neq  id_i)$ 
   {\bf then} discard the message
      (do not execute lines~\ref{AG-32}-\ref{AG-33}) {\bf end if.}\\

\line{AG-32}  \>   
$\dtwocol_i \leftarrow \dtwocol_i  \cup  \donecol$;  
$\donecol_i \leftarrow \donecol_i  \cup  color$;\\

\line{AG-33}  \>  \ $color_i\leftarrow $ 
 the first  color in $0, 1..$ which is not in
  $\donecol_i \cup \dtwocol_i$; $sender_i \leftarrow sender$; 
$state_i \leftarrow 4$.
 
\\~\\


{\bf when} 
{\sc corrected\_color}$(dest1, dest2, sender,color)$
 {\bf is received} {\bf do} \\

\line{AG-34} \> {\bf if}  $(dest1 \neq id_i) \wedge (dest1 \neq -1)   $  

 {\bf then} \= ${\sf delete}$  the last color added to $\donecol_i$; 
$\donecol_i \leftarrow \donecol_i \cup \{color\}$  {\bf end if}; \\

\line{AG-35} \> {\bf if}  $(dest1 \neq id_i) \wedge (dest1 = -1)$  

 {\bf then} \= ${\sf delete}$  the last color added to $\dtwocol_i$; 
$\dtwocol_i \leftarrow \dtwocol_i \cup \{color\}$  {\bf end if}; \\

\line{AG-36} \> 

{\bf if}  $(dest2 = id_i)$  

 {\bf then} \=  $state_i \leftarrow 6$; $corrected\_cl_i \leftarrow color$  
{\bf end if};\\

\line{AG-37} \> {\bf if} \= $(dest1 = -1) \wedge (dest2 = -1) 
\wedge (color_i = color) $ 
{\bf then} \=  $state_i \leftarrow 5$ {\bf end if}. \\~\\


{\bf when} 
{\sc resume\_coloring}$(dest, sender)$ {\bf is received} {\bf do} \\

\line{AG-38} \>
{\bf if} $(dest \neq  id_i)$ 
   {\bf then} discard the message
      (do not execute lines~\ref{AG-39}-\ref{AG-40}) {\bf end if};\\

 \line{AG-39} \> $parent_i \leftarrow sender$;\\

\line{AG-40} \> {\bf if} \=   $(to\_color_i \neq \emptyset)$ {\bf then} \=  
  
 $state_i \leftarrow 2$  {\bf else}   $state_i \leftarrow 3$ {\bf end if}.

\end{tabbing}
\normalsize
\end{minipage}
}
\caption{Sequential distance-2 coloring for an arbitrary graph  
(code for $p_i$)}
\label{fig:coloring-of-an-arbitrary-graph}
}
\end{algorithm}

\begin{figure}[ht]
{\centering\includegraphics[height=5cm]{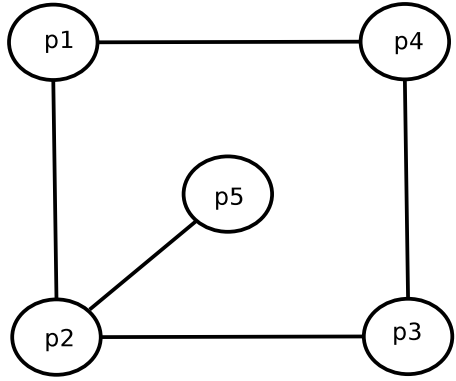} 
    \caption{A 5-process arbitrary network}
\label{arbitrary-graph}
}
\end{figure}

Considering the network of Figure~\ref{arbitrary-graph}, An example of
execution of Algorithm~\ref{fig:coloring-of-an-arbitrary-graph} is
depicted in Table~\ref{fig:table-example}. Due to space restriction we
abbreviate the following: \\

\begin{itemize}

\item br =   ${\sf broadcast}$ operation,

\item $d1_i$ = $\donecol_i$,
\item $d2_i$ = $\dtwocol_i$,
\item CL($p_a$,$p_b$,c,z,S) = \ccolor($p_a$,$p_b$,c,z,S),
\item  CRL($p_a$,$p_b$,c,S) = {\sc correct}($p_a$,$p_b$,c,z,S),
\item  CR\_CL($p_a$,$p_b$,c,S) =  {\sc corrected\_cl}($p_a$,$p_b$,c,S), 
\item   RSM\_CL($p_a$,$p_b$) = {\sc resume\_cl}($p_a$,$p_b$),
\item $s_i=state_i$.

\end{itemize}

Process $p_1$ receives the message {\sc start}$()$ at round $-1$. It
broadcasts the message \ccolor($id_1$, $id_1$, -1, 0, $\emptyset$) to
itself. It receives this message at round $0$.  It updates its local
variables $d2_1$ at line 02 (abbreviated as L2), $d1_1$, $s_1$ and
$color_1$ at line 06 (abbreviated as L6). This appears in the first
row of the table where the value of $\CLOCK$ is $0$. Then, when
$\CLOCK$ progresses to $1$, $p_1$ has $state_1=2$, therefore, it
broadcasts the message {\sc color}$()$ with appropriate parameters,
where it proposes a color that is not in $d2_1\cup d1_1$ to each of
its neighbors, and subsequently enters $state_1= 0$ (L16). This
appears in the second row of the table where the value of $\CLOCK$ is
$1$.  When $p_2$ and $p_4$ receive this message at round $2$, they
execute the associated processing at L2 and L6 for $p_2$ and at L10
for $p_4$ . So, in this round $p_2$ updates its local variables $d2_2$
at L2, $d1_2$, $s_2$ and $color_2$ at L6.  $p_4$ updates its local
variables $d2_4$ at L2, $d1_4$, at L10. This appears in the third row
of the table where $\CLOCK$ is $2$. In round 3, $p_2$ has $state_2=2$,
so it broadcasts the message {\sc color}$()$ with appropriate
parameters (L16) and gets $state_2= 0$ (L16), Etc.

In round 6, $p_4$ finds that the color proposed by its neighbor $p_3$
(color 0 proposed in round 5) is in $d1_4$, so it ''refuses'' this
color and gets $state_4= 1$ (L4).  In round 7, $p_4$ gets a color
which is not in $d2_4\cup d1_4$ (L11) and broadcasts the message
{\sc correct}($id_3$, $id_4$, 2, 0, $\lbrace 0\rbrace$). In round 8, $p_3$
updates $d1_3$ and $d2_3$ and gets a new color (L31). Its state will
allow it to broadcast the message {\sc corrected\_cl}$()$ (round 11). The
broadcast of this message will trigger the broadcast of the message
{\sc resume\_cl}$()$ (round 15) to resume the coloring as described before.

\begin{table}[h]
\scriptsize
  \begin{tabular}{|p{0.66cm}|p{2.7cm}|p{3.4cm}|p{2.95 cm}|p{2.46cm}|p{1.93cm}|}%
 \hline
\theadfont\diagbox[width= 3.3em]{ \hspace{-5 pt}clock}{$p_i$}
&     \hspace{-5 pt}    $p_1$ & $p_2$ & $p_3$ & $p_4$ & $p_5$   \\\hline
0 &\hspace{-12.5 pt}\,$\,\,\,\,\,d2_1$\,=\,\,$\lbrace\rbrace$\,(L2),\hspace{ 0pt}$d1_1$\,=\,\,$\lbrace-1\rbrace$ & & & & \\
&\hspace{-7 pt} $s_1 = 2$, $color_1 = 0$ (L6) &  & & &
\\\hline 
 1& \hspace{-8 pt} br CL($id_2$,$id_1$,0,1,$\lbrace -1 \rbrace$)& & & & \\
&\hspace{-5 pt}$s_1 = 0$ (L16) &  & & & \\\hline
 
  2& & \hspace{-4 pt}  $d2_2=\lbrace-1\rbrace$ (L2),$d1_2 = \lbrace 0 \rbrace$ $s_2 = 2$, $color_2 = 1$ (L6)& & \hspace{-8 pt} $d2_4=\lbrace 1\rbrace$,$d1_4=\lbrace0\rbrace$ (L10)& \\\hline
   3& & br CL($id_3$,$id_2$, 1, 2, $\lbrace 0\rbrace$) & & &  \\
& & $s_2 = 0$ (L16) & & &\\\hline

    4&\hspace{-5 pt} $d2_1 = \lbrace -1, 2 \rbrace$, & & $d2_ =\lbrace0\rbrace$ (L2),$d1_3 = \lbrace1\rbrace$ & & $d2_5=\lbrace0,2\rbrace$ \\
&\hspace{-5 pt}$d1_1 = \lbrace 1 \rbrace$ (L10) &  & $s_3 = 2$, $color_3 = 2$ (L6) & & \hspace{-5 pt} $d1_5 = \lbrace 1 \rbrace$ (L10) \\\hline

 5&  & & br CL($id_4$,$id_3$, 2, 0, $\lbrace 1\rbrace$) & & \\
& &  &  $s_3 = 0$ (L16) & &  \\\hline 

 6 &  & $d2_2$\,=\,\,$\lbrace-1,0\rbrace$, \hspace{25 pt} $d1_2 = \lbrace 0, 2 \rbrace$ (L10)   & & $d2_4 = \lbrace 1 \rbrace$ (L2), \hspace{20 pt}$s_4 = 1$ (L4) &   \\\hline 

 7 & & & &  $color_4=2$ (L11), \hspace{8 pt} br CR($id_3$,$id_4$,2,$\lbrace 0\rbrace$) & \\
& &  &   & $s_4 = 0$ (L12)&  \\\hline 

8 & &  & $d2_3$\,=\,\,$\lbrace 0 \rbrace$, \hspace{25 pt} $d1_3$\,=\,\,$\lbrace 1,2\rbrace$ (L30) &   & \\
& &  &  $color_3 = 3, s_3 = 4$ (L31) &  &  \\\hline 

9 & &  & \hspace{-4 pt} br CR\_CL($id_4$,$id_2$,$id_3$,3) &   & \\
& &  & $s_3 = 0$ (L19) &  &  \\\hline 

10 & & $d1_2$\,=\,\,$\lbrace 0,3 \rbrace$\,(L32), \hspace{20 pt}$s_2 = 6$ (L34) &  &   &  \\\hline 

11 & &  br CR\_CL($-1$,$-1$,$p_2$,3) &  &   & \\
& & $s_2 = 0$ (L21) &   &  &  \\\hline

12 & $d2_1 = \lbrace -1, 3\rbrace$ (L33) & &  &    &    \\\hline

13 & & & $s_3 = 5$ (l35)  &    & \hspace{-7.9 pt}$\,\,d2_5$\,=\,\,$\lbrace0,3\rbrace$ (L32) 
 \\\hline

14 & & & br RSM\_CL($id_4$,  $id_3$)  &    &  \\
& & & $s_3 = 0$ (L20)   &  &  \\\hline

15 & & &  &  $s_4 = 3$ (L38)   &   \\\hline

16 & & &  & br TERM($id_3$,  $id_4$)  &  \\
& & &   & $s_4 = 0$ (L18) &  \\\hline

17 & & & $s_3 = 3$ (L38) &    &   \\\hline

18 & & & br TERM($id_2$,  $id_3$) &   &  \\
& & &  $s_3 = 0$ (L18) &  &  \\\hline

19 & &  $s_2 = 2$ (L27)  &  &   &  \\
& & $d1_2 = \lbrace 0,3\rbrace$ (L26) & &  &  \\\hline

20 & &  br\,CL($id_5$,$id_2$,1,2,$\lbrace 0,3\rbrace$)\,(L16)  &  &   &  \\
& & $s_2 = 0$ (L16) & &  &  \\\hline

21 & $d2_1$\,=\,\,$\lbrace -1,3,2\rbrace$, $d1_1 = \lbrace 0, 1 \rbrace$ (L10) &  & $d2_3$\,=\,\,$\hspace{-3 pt}\lbrace 0,2\rbrace$\,,\, \hspace{20 pt} $d1_3 = \lbrace 2,1\rbrace$ (L10) &   & \hspace{-5 pt}$d2_5$\,=\,\,$\lbrace 0, 3\rbrace$ \hspace{-1 pt} $s_5 = 3$ (L07)  \\\hline

22 & &  &  &   & \hspace{-0.27 cm }  br\,TERM($id_2$,$id_5$) \\
& & & &  & \hspace{-8 pt} $s_5 = 0$ (L18) \\\hline

23 & &  $s_2 = 3$ (L25) &  &   &  \\
& & & &  & \\\hline

24 & &  br TERM($id_1$,  $id_2$)  &  &   &  \\
& & $s_2 = 0$ (L18) & &  & \\\hline

25 & end algorithm (L25) &    &  &   &  \\
& & & &  & \\\hline

\end{tabular}
\caption{An execution of Algorithm~\ref{fig:coloring-of-an-arbitrary-graph}
on the network of Figure~\ref{arbitrary-graph}}
\label{fig:table-example}
\end{table}

\end{document}